\documentclass[aps,prl,10pt,reprint,superscriptaddress]{revtex4-1}

\usepackage{amsmath,amsfonts,amssymb}
\usepackage{xcolor,graphicx}
\usepackage{braket}
\usepackage[pdftex]{hyperref}
\usepackage{csquotes}
\hypersetup{
	colorlinks = true,
	linkcolor = blue,
	anchorcolor = blue,
	citecolor = red,
	filecolor = blue,
	urlcolor = blue}

\begin{document}
\title{Cauchy-Riemann beams in GRIN media}

\author{I. Ramos-Prieto}
\email[e-mail:\,]{iran@inaoep.mx}
\affiliation{Instituto Nacional de Astrofísica Óptica y Electrónica, Calle Luis Enrique Erro No. 1\\ Santa María Tonantzintla, Puebla, 72840, Mexico}

\author{D. Sánchez-de-la-Llave}
\affiliation{Instituto Nacional de Astrofísica Óptica y Electrónica, Calle Luis Enrique Erro No. 1\\ Santa María Tonantzintla, Puebla, 72840, Mexico}

\author{U. Ruíz}
\affiliation{Instituto Nacional de Astrofísica Óptica y Electrónica, Calle Luis Enrique Erro No. 1\\ Santa María Tonantzintla, Puebla, 72840, Mexico}

\author{V. Arriz\'on}
\affiliation{Instituto Nacional de Astrofísica Óptica y Electrónica, Calle Luis Enrique Erro No. 1\\ Santa María Tonantzintla, Puebla, 72840, Mexico}

\author{F. Soto-Eguibar}
\affiliation{Instituto Nacional de Astrofísica Óptica y Electrónica, Calle Luis Enrique Erro No. 1\\ Santa María Tonantzintla, Puebla, 72840, Mexico}

\author{H. M. Moya-Cessa}
\affiliation{Instituto Nacional de Astrofísica Óptica y Electrónica, Calle Luis Enrique Erro No. 1\\ Santa María Tonantzintla, Puebla, 72840, Mexico}

\date{\today}

\begin{abstract}
We investigate the propagation characteristics of Cauchy-Riemann beams in gradient-index media. Our study reveals two key findings: a) the preservation of their form during propagation, and b) surprisingly, the feasibility of obtaining the fractional Fourier transform for any arbitrary entire function. We provide an explicit and straightforward expression for this transform. Additionally, our results contribute to a deeper understanding of the behavior of Cauchy-Riemann beams in complex media, offering insights that may have implications for various applications in optics and wave propagation.
\end{abstract}
\maketitle
\textbf{Introduction.} Propagation of light beams is an important process in applied and theoretical optics. Over the years, special attention has been given to light beams that preserve their intensity profile during propagation. Several non-diffracting beams, such as Bessel~\cite{Durnin_1987,McGloin_2005,Fahrbach_2010,Gori_1987}, Mathieu~\cite{Gutierrez_2000}, and Weber beams~\cite{Bandres_2004}, keep their transverse intensity profiles without a scale change. However, the practical implementation of those beams requires including a finite envelope~\cite{Gori_1987,khonina_1999,Siviloglou_2007,Bandres_2007,Khonina_2011}, and the invariant structure is only preserved in a finite region. There are also optical fields that preserve their shape in an unbounded propagation range, the so-called scaled propagation invariant (SPI) beams~\cite{Arrizon_2018}, but this is achieved at the expense of a re-scaling in their transverse extensions. Hermite-Gaussian and Laguerre-Gaussian beams are good examples of these optical fields~\cite{Siegman_1986}. There exist methods intended to extend the existence domain of Bessel beams~\cite{Jaroszewicz_1998,Aruga_1999,Belyi_2010,Ismail_2012}. These beams need to be modified in a way that a re-scaling in their transverse widths is also present. Recently \cite{Arrizon_2018}, we have produced an approximate method that, by using a positive lens,  any optical field may be converted into an SPI beam. Propagation in quadratic gradient-index (GRIN) media may be viewed as a beam splitter \cite{Soto-Eguibar:14}.

In \cite{CRB}, it was shown that given an entire function $f(x+iy)$, the field
\begin{equation}\label{0010}
	E(x,y,z)=\frac{\exp{\left[-\frac{g(x^2+y^2)}{w(z)}\right]}}{w(z)} f\left(\frac{x+iy}{w(z)}\right),
\end{equation}
with $g$ an arbitrary complex constant and $w(z)=1+i2gz$, is a solution of the paraxial wave equation, satisfying the initial condition $E\left(x,y,0\right)=\exp{\left[-g \left(x^2+y^2\right)\right]} f(x+iy)$.  In light of $f(x+iy)$ satisfying the Cauchy-Riemann conditions, we have opted to name these fields Cauchy-Riemann beams, a decision of paramount significance for Eq.~\eqref{0010} to qualify as a solution to the paraxial equation. Through the appropriate choice of the analytic function $f(x+iy)$, the Cauchy-Riemann beams, as denoted in Eq.~\eqref{0010}, are transversely square integrable. Furthermore, we showed, theoretically and experimentally, that these beams rotate, and we explained such a rotation employing the Bohm potential.

In this letter, we expand upon the previously mentioned research within the realm of fields propagating in GRIN media. We delve into both quadratic and linear GRIN media scenarios. Firstly, the quadratic GRIN media case holds paramount significance, as the propagator associated with the paraxial wave equation also happens to be the operator for the two-dimensional fractional Fourier transform~\cite{Namias_1980,Mendlovic:93}. This discovery empowers us to formulate a general expression for the fractional Fourier transform of any entire function. This groundbreaking result, hitherto unknown, bears profound importance in our research. Secondly, we present the exact analytical solution for linear GRIN media, expressed in terms of any entire function $f(x+iy)$. In both cases, we take advantage of the fact that $\nabla_\perp^2 f(x+iy)=0$, to fully utilize the established techniques of quantum optics.

\textbf{Quadratic GRIN media.} Propagation of light in a quadratic GRIN medium obeys the paraxial wave equation
\begin{equation}\label{ecschcua}
	i\frac{\partial E(x,y,z)}{\partial z}=-\frac{\nabla_\perp^2}{2} E(x,y,z)+\frac{\eta^2}{2}\left(x^2+y^2\right) E(x,y,z),
\end{equation}
where $\nabla_\perp^2$ is the transversal Laplacian, and $\eta$ is a real constant that characterizes the medium. Comparing the previous paraxial equation with the Schrödinger equation of two decoupled harmonic oscillators, it becomes evident that they are equivalent when we identify the time $t$ with the propagation distance $z$. In the subsequent analysis, we adopt an approach involving operators, an approach less commonly employed in physical optics~\cite{Stoler_1980}. Nevertheless, this approach proves advantageous for obtaining the propagator or evolution operator for two decoupled oscillators. In \cite{ASENJO2021126947}, it is demonstrated that the propagator associated with Eq.~\eqref{ecschcua} is
\begin{equation}\label{U_z}
	\begin{split}
		&\hat{U}(z)= \exp\left[-i\eta^2 \alpha(z)\sum_{q=x,y}q^2\right]\\
		&\times\exp\left[-i\beta(z)\sum_{q = x,y}\left(q\hat{p}_q+\hat{p}_qq\right)\right]\exp\left[-i\alpha(z)\sum_{q = x,y} \hat{p}_q^2\right],
	\end{split}
\end{equation}
with $\alpha\left(z\right)=\frac{\tan\left(\eta z \right)}{2\eta}$, $\beta\left(z\right)=\ln\left[\cos\left(\eta z \right)  \right]/2$, and where $\hat{p}_x=-i\partial/\partial x$ and $\hat{p}_y=-i\partial/\partial y$ are the standard momentum operators.  The commutation relations within the system are expressed as $[x,\hat{p}_x]= [y,\hat{p}_y]= i$ and $[x,y]=[x,\hat{p}_y]= [y,\hat{p}_x]=[\hat{p}_x,\hat{p}_y]=0$. As we will see in the next section, this propagator is the operator of the two-dimensional fractional Fourier transform. By imposing the initial condition $E(x, y, 0) = \exp\left[-g(x^2 + y^2)\right]\mathcal{E}(x+iy)$, where $g\in\mathbb{C}$ and $\mathcal{E}(x+iy)$ represents an analytic function, the solution to Eq.~\eqref{ecschcua} can be determined. To accomplish this task, it is crucial to recognize that the operators involved in the propagator $\hat{U}(z)$, as given by Eq.~\eqref{U_z}, namely $q^2$, $\hat{p}^2_q$, and $q\hat{p}_q+\hat{p}_qq$, are closed under commutation (with $q=x,y$). Furthermore, the operators involved in the propagator correspond to translations and squeezings of the transverse coordinates. Importantly, this consideration is made while taking into account that $\exp\left[-i\alpha(z)\sum_{q=x,y}\hat{p}_q^2\right]\mathcal{E}(x+iy) =\mathcal{E}(x+iy)$. Through the application of common techniques in quantum optics and a lengthy yet straightforward calculation, we have determined that the solution is as follows:
\begin{equation}\label{solcuadr}
	\begin{split}
		E(x,y,z)&=\frac{\exp\left[-g(x^2+y^2)
				\frac{1+i\frac{\eta}{2g}\tan(\eta z)}{1+i\frac{2g}{\eta}\tan(\eta z)}\right]}
		{\cos(\eta z)+i\frac{2g}{\eta}\sin(\eta z)}\\&\times
		\mathcal{E}\left(\frac{x+iy}{\cos(\eta z)+i\frac{2g}{\eta}\sin(\eta z)}\right).
	\end{split}
\end{equation}
By directly substituting into the paraxial equation, one can confirm the validity of the proposed solution. Furthermore, the verification of the initial condition $E(x,y,0) = \exp\left[-g(x^2 + y^2)\right]\mathcal{E}(x+iy)$ is easily achieved. This analysis of the solution, as part of our ongoing research, constitutes one of our primary contributions in this field.
\begin{figure}
	\centering
	\includegraphics[width=\linewidth]{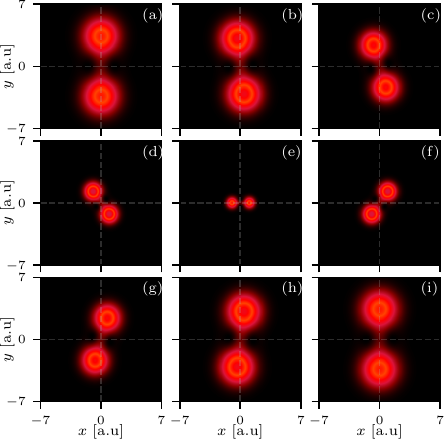}
	\caption{Intensity propagation of the field described by Eq.~\eqref{solcuadr} with $\mathcal{E}(x+iy)=\cos\left[\mu\left(x+iy\right)\right]$. Intensity is plotted at varying distances $z$, ranging from $0$ to $\pi$ in increments of $\pi/8$. For the sake of simplicity and without loss of generality, we have considered the parameters $\{g, \mu, \eta\} =\left\{0.15,1,1\right\}$ to characterize the simulations, influencing the observed evolution of intensity.}
	\label{fig_1}
\end{figure}
Illustrating a key contribution in this manuscript through Eq.~\eqref{solcuadr}, we examine the entire function with the specific choice $\mathcal{E}(x+iy)=\cos\left[\mu\left(x+iy\right)\right]$, as shown in Fig.~\ref{fig_1}. Noteworthy is the evolution of the initial condition observed between subfigures (a) and (e) up to a distance of $\pi/2$, followed by a reversed evolution pattern between figures (e) and (i) across distances from $\pi/2$ to $\pi$. As we will see shortly, this distinctive behavior is inherent to fractional orders in the two-dimensional fractional Fourier transform.

\textbf{The fractional Fourier transform of an entire function.} The computation of a two-dimensional fractional Fourier transform is indeed a challenging task. Here, to the best of our knowledge, we present, for the first time, an expression for the fractional Fourier transform of an arbitrary entire function.\\
As mentioned earlier, the propagator $\hat{U}(z)$ in Eq.~\eqref{U_z} serves as the operator for the two-dimensional fractional Fourier transform. To facilitate understanding, and for the sake of simplicity, let us treat the propagator as a function of one of the transverse coordinates denoted by $q=x, y$, and we make $\eta = 1$. Thus, the propagator is formally expressed as $\hat{U}_q(z) = \exp[-i\alpha(z)q^2]\exp[-i\beta(z)(q\hat{p}_q+\hat{p}_qq)]\exp[-i\alpha(z)\hat{p}_q^2]$. Given these considerations, our objective is to establish that the expression
\begin{equation}\label{FRFT}
\mathcal{F}_{z}^{1D}\left\{s\right\}(q)=\exp\left[i\frac{z}{2}\right]\hat{U}_q(z)s(q)
\end{equation}
is the fractional Fourier transform. To achieve this, we apply it to a function $s(q)$, which we express in terms of its inverse Fourier transform, defined as
\begin{equation}
s(q) = \frac{1}{\sqrt{2\pi}} \int\limits_{-\infty}^{\infty} du \, S(u) \exp(-iuq),
\end{equation}
with the Fourier transform defined as
\begin{equation}\label{S_u}
S(u) = \frac{1}{\sqrt{2\pi}} \int\limits_{-\infty}^{\infty} d\xi \, s(\xi) \exp(i\xi u).
\end{equation}
We introduce the first operator given in Eq.~\eqref{U_z} (with $\eta = 1$), i.e., $\exp[-i\alpha(z)\hat{p}_q^2]$, inside the integral, and we use the fact that $\exp\left(-iuq\right)$ is an eigenfunction of the operator $\hat{p}_q$ with eigenvalue $-u$, to obtain
\begin{equation}
\begin{split}
&\mathcal{F}_z^{1D}\left\{s\right\}(q) = \exp\left(i\frac{z}{2}\right) \exp\left(-i\frac{\tan z}{2}q^2\right)\\
&\times \exp\left[-i\frac{\ln(\cos z)}{2} \left(q\hat{p}_q + \hat{p}_qq\right)\right]\\
&\times \frac{1}{\sqrt{2\pi}} \int\limits_{-\infty}^{\infty} du \, S(u) \exp\left(-i\frac{\tan \alpha}{2}u^2\right) \exp(-iqu).
\end{split}
\end{equation}
Considering that
\begin{equation}
\begin{split}
\exp[-i\beta(z)(q\hat{p_q}+\hat{p}_qq)]&\exp(-iqu)\exp[i\beta(z)(q\hat{p_q}+\hat{p}_qq)]\\
&=  \exp(-iqu\sec z),\\
\exp[-i\beta(z)(q\hat{p_q}+\hat{p}_qq)] 1&= \sqrt{\sec z},
\end{split}
\end{equation}
and going back to the original function $s(q)$ written in terms of its Fourier transform, Eq.~\eqref{S_u}, we obtain the one-dimensional fractional Fourier transform~\cite{Namias_1980}
\begin{equation}
	\begin{split}
		\mathcal{F}_z^{1D}\left\{s\right\}(q)&=\sqrt{\frac{1-i\cot z}{2\pi}}\exp\left(i\frac{\cot z}{2}q^2\right)\\
		&\times\int\limits_{-\infty}^{\infty}d\xi{s}(\xi)
		\exp\left(i \frac{\cot z}{2}\xi^2-iq\xi\csc z\right).
	\end{split}
\end{equation}
A generalization of the two-dimensional fractional Fourier transform for a function of two variables, namely $s(\xi,\chi)$, can be demonstrated by following the aforementioned steps and considering the complete propagator $\hat{U}(z)$, such that
\begin{equation}
	\begin{split}
		&\mathcal{F}_z^{2D}\left\{s\right\}(x,y)=\frac{1-i\cot z}{2\pi}\exp\left[i\frac{\cot z}{2}\left(x^2+y^2\right)\right]\\
		&\times\int\limits_{-\infty}^{\infty}\int\limits_{-\infty}^{\infty}d\xi d\chi{s}(\xi,\chi)\\
		&\times\exp\left[i \frac{\cot z}{2}\left(\xi^2+\chi^2\right)-i\left(x\xi+y\chi\right)\csc z\right].
	\end{split}
\end{equation}
However, given that $\exp[-i\alpha(z)(\hat{p}_x^2+\hat{p}_y^2)]\mathcal{E}(x+iy) = \mathcal{E}(x+iy)$, where $s(\xi,\chi)\rightarrow\mathcal{E}(x+iy)$ satisfies the Laplace equation with eigenvalue zero, an outcome derived from the Cauchy-Riemann equations, it is evident that the two-dimensional fractional Fourier transform of an entire function is
\begin{equation}\label{Solution}
	\begin{split}
		\mathcal{F}_z^{2D}\left\{\mathcal{E}\right\}(x,y)&=\exp[iz]\hat{U}_x(z)\hat{U}_y(z)\mathcal{E}(x+iy)\\
		&=\frac{\exp[iz]}{\cos z}\exp\left[-i\frac{\tan z}{2}(x^2+y^2)\right]\\
		&\times\mathcal{E}\left(\frac{x+iy}{\cos z}\right).
	\end{split}
\end{equation}
where we have used the propagator defined by Eq~\eqref{U_z}, $\hat{U}(z)=\hat{U}_x(z)\hat{U}_y(z)$ and  (with $\eta = 1$). This result is one of the main findings of this manuscript, as the two-dimensional fractional Fourier transform can be easily obtained for any entire function $\mathcal{E}$.

\begin{figure}
	\centering
	\includegraphics[width=\linewidth]{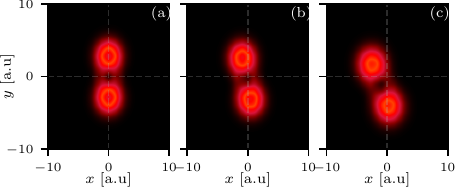}
	\caption{Propagation of the intensity of the field given in \eqref{sollineal}, when $\mathcal{E}(x+iy)=J_1\left[\mu\left(x+iy\right)\right]$. The intensity is plot at $z$, from $0.00$ to $\pi/2$ in increments of $\pi/4$. The parameters used are: $\left\{g,\mu,\eta_x,\eta_y\right\}=\left\{0.15,1,1,1\right\}$.}
	\label{fig_2}
\end{figure}

\textbf{ Linear GRIN media.} For a lineal GRIN medium, the propagation of light satisfies the paraxial wave equation
\begin{equation}\label{pareqlin}
	i\frac{\partial E(x,y,z)}{\partial z}=-\frac{\nabla_\perp^2}{2} E(x,y,z)+\left(\eta_x x+\eta_y y\right) E(x,y,z),
\end{equation}
where $\eta_x$ and $\eta_y$ are real constants that characterize the medium. Again, employing techniques from quantum optics operators, it can be demonstrated that the propagator, in this case, is \cite{Soto-Eguibar_2015}
\begin{equation}
	\begin{split}
		\hat{U}(z)&= \exp\left[-i\frac{(\eta_x^2+\eta_y^2)z^3}{6}\right]\exp[-i(\eta_xx+\eta_yy)z]\\&\times
		\exp\left[i\frac{\eta_xz}{2}\hat{p}_x+i\frac{\eta_yz}{2}\hat{p}_y\right]\exp\left[-i\frac{z}{2}(\hat{p}_x^2+\hat{p}_y^2)\right].
	\end{split}
\end{equation}
Imposing the same initial conditions that in the quadratic case, i.e. $E(x,y,0)=\exp[-g(x^2+y^2)]\mathcal{E}(x+iy)$, we found after a cumbersome calculation that the solution to this initial value problem is
\begin{equation}\label{sollineal}
	\begin{split}
		&E(x,y,z)= \frac{\exp\left[-i\frac{z^3}{6} \left(\eta_x^2+\eta_y^2\right) \right]}{{w(z)}}\exp[-i z \left(\eta_xx+\eta_yy\right)]\\&\times\exp\left\{ -\frac{g}{w(z)}\left[\left( x+\frac{\eta_xz^2}{2}\right) ^2+\left( y+\frac{\eta_yz^2}{2}\right) ^2\right]\right\}\\
		&\times \mathcal{E}\left(\frac{x+\dfrac{\eta_x z^2}{2}+i\left(y+\dfrac{\eta_y z^2}{2}\right)}{w(z)}\right),
	\end{split}
\end{equation}
where $w(z)=1+2igz$, as in the free propagation case.\\
By direct substitution of Eq. \eqref{sollineal} in the paraxial equation (\ref{pareqlin}), it can be verified that it is indeed a solution, and it is also trivial to check that the initial condition is satisfied.\\
In Fig.~\ref{fig_2}, we plot the solution \eqref{sollineal} of the paraxial wave equation \eqref{pareqlin}, when the entire function is $\mathcal{E}(x+iy)=J_1\left[\mu\left(x+iy\right)\right]$ for a linear GRIN medium, where $J_1\left(\zeta\right)$ is the Bessel function with $n=1$.

\textbf{Conclusions.} In conclusion, our investigation into the propagation of Cauchy-Riemann beams in gradient-index media has yielded significant insights. The observed preservation of their form throughout propagation suggests robust characteristics that could be advantageous for applications requiring beam stability. Moreover, the unexpected discovery of obtaining the fractional Fourier transform for arbitrary entire functions introduces a novel and efficient approach to manipulating optical signals in complex media. These findings not only deepen our understanding of the behavior of Cauchy-Riemann beams but also open avenues for innovative applications in optics and wave propagation. The explicit and simple expression for the fractional Fourier transform adds a practical dimension to the theoretical framework, facilitating its potential incorporation into diverse optical systems. Overall, our study contributes valuable knowledge to the field, with implications for advancing technologies reliant on controlled beam propagation in gradient-index media.

%
\end{document}